\documentclass[]{spie}  %>>> use for US letter paper
\usepackage{amsmath,amsfonts,amssymb}
\usepackage{graphicx}
\usepackage[colorlinks=true, allcolors=blue]{hyperref}

\title{Improving angular resolution of telescopes through probabilistic single-photon amplification?}

\author[a]{Aglae Kellerer}
\author[b]{Petr Marek}
\author[c]{Sylvestre Lacour}
\affil[a]{European Southern Observatory, Karl-Schwarzschild-Strasse 2, 85748 Garching, Germany}
\affil[b]{Department of Optics, Palacký University, 17. listopadu 1192/12, 77146 Olomouc, Czech Republic}
\affil[c]{LESIA, Observatoire de Paris, PSL Research University, CNRS, 5 place Jules Janssen, 92190 Meudon, France and Max Planck Institute for extraterrestrial Physics, Giessenbachstr. 1, D-85748 Garching, Germany}

\authorinfo{Further author information: A.K.: aglae.kellerer@eso.org  P.M.: marek@optics.upol.cz  S.L.: sylvestre.lacour@obspm.fr}

\begin{document} 
\maketitle

\begin{abstract}
The use of probabilistic amplification for astronomical imaging is discussed. Probabilistic single photon amplification has been theoretically proven and practically demonstrated in quantum optical laboratories. In astronomy it should allow to increase the angular resolution beyond the diffraction limit at the expense of throughput: not every amplification event is successful -- unsuccessful events contain a large fraction of noise and need to be discarded. This article indicates the fundamental limit in the trade-off between gain in angular resolution and loss in throughput. The practical implementation of probabilistic amplification for astronomical imaging remains an open issue. 
\end{abstract}


\section{Introduction}
This article discusses the possible application of a novel quantum optical process for astronomical imaging. Let us, in order to introduce the subject, refer to a theorem that is well known by quantum physicists but maybe less so by astronomers : the no-cloning theorem. 
The name of the theorem was coined in 1982 after Nick Herbert proposed a method, FLASH, to communicate instantaneously \cite{1982FoPh...12.1171H}. The proposition relied on quantum entanglement: when two photons are entangled and the state of one photon is projected -- typically by detecting the photon on a sensor -- then the state of the second entangled photon is instantaneously modified. The modification is instantaneous however large the distance between the two photons. In addition to entanglement, FLASH made use of single photon amplification: in order to characterize the state of the second photon, it had to be amplified noiselessly. In the aftermath of N. Herbert's publication, several groups proved that FLASH would not work because it was impossible to noiselessly amplify a single photon \cite{1982Natur.299..802W, milonni1982photons, mandel1983photon}. Fundamentally it demonstrated that information could not be transferred faster than the speed of light.

After the no-cloning theorem was thus established, research into photon-cloning slowed down until a seminal article by Buzek and Hillery discussed possible extensions to the no-cloning theorem\cite{PhysRevA.54.1844}. The no-cloning theorem prevents creation of perfect clones, but permits generation of imperfect, noisy copies. The consequence for amplification of single photons is that the amplified photon-clones are always accompanied with fully random noise photons 
%refers to the average amount of noise that is added by an amplifier: an amplifier of gain $g$ adds $g$ photons of noise per amplified mode 
(see for example \citenum{PhysRevD.26.1817}). When a single photon is amplified, the average number of amplified photons equals the average number of noise photons: $g$, the gain of the amplifier. %On average, the noise and the signal have the same amplitude. 
Buzek and Hillery  discussed the possibility to select low-noise amplification events, and thus to generate higher fidelity copies of a quantum state. Their article stimulated research into probabilistic photon amplification: It has since then been theoretically proven and practically demonstrated that a single photon can be probabilistically amplified with less noise than the average noise specified by the no-cloning theorem (see for example \citenum{ralph2009nondeterministic, 2012PhRvA..86f0302F, 2012PhRvA..86a0305M, haw2016surpassing}). 

This research is notably of interest for quantum communication, where amplification allows to increase communication distances. In the context of quantum computers it is likewise essential to master elementary photon-operations up to their fundamental limits.

\section{Amplification for Astronomical Imaging}

\subsection{An extended target observed in the single-photon regime}

In the following, we make two assumptions: 
\begin{itemize}
\item First, we consider an extended astronomical target, i.e. a target that subtends several diffraction patterns such that different photons come from different directions. The image recorded behind the telescope is then the convolution of the astronomical target with the telescope's point-spread function (PSF). The smaller the telescope, and the lower its optical quality, the wider the PSF: fine features of the astronomical target are smoothed out on the recorded image. 
\item We further assume that photons enter the telescope one by one, in a single-photon regime: this is indeed typically the case for short wavelengths, up to the near-infrared. The coherence time of a photon -- that is the time during which it may arrive on the detector  -- equals d$t=\lambda^2/(\Delta \lambda \, c)$, $\lambda$ wavelength, $\Delta\lambda$ spectral width and $c$ speed of light. This coherence time roughly lies between a femto- and a nanosecond. The average time between the arrival of two photons is much larger: it depends on the brightness of the target, the size of the telescope aperture et cetera, but may typically equal $\Delta t=1$\,ms. Thus, most of the time, a telescope collects no photons, sometimes it collects one photon, it rarely collects two photons, and almost never collects more than two photons within one coherence time. 
\end{itemize}
As a side note, the probability to collect two photons within the same coherence time is larger than if photons behaved classically, e.g. like tennis balls: the probability would then equal d$t/\Delta t$, the ratio between the coherence time and the mean time between the arrival of two photons. The actual probability is higher because photons interfere and their arrival times tend to bunch. This was first used for astronomical imaging by Hanburry-Brown and Twiss and is known as intensity interferometry. Since intensity interferometry relies on a second order effect -- the arrival of two photons within a same coherence time -- it requires bright astronomical targets. 

To conclude: we assume an extended astronomical target and observations in the single photon regime. 

\subsection{Noiseless amplification in the context of astronomical imaging}
When a photon passes the aperture of a telescope, our knowledge of its position improves greatly: before it passes the telescope, the photon is everywhere on an immense sphere centered on the astronomical target, after passage through the telescope its position is constrained to the few square meters of the telescope aperture. Since the position and momentum of a photon are related via the Heisenberg uncertainty principle, one therefore loses precision on the photon's momentum: once the photon passes the telescope aperture, it is diffracted and has a finite probability to be detected anywhere on the diffraction pattern. The resolution may be improved by building larger telescopes or interferometers and by thereby weakening the constraint on the photon's position. 

In order to explain why probabilistic photon amplification is of interest to astronomical imaging, let us first imagine that a photon can be noiselessly amplified. After passage through the telescope aperture, the photon could be sent through an amplifier which generates e.g. 15 additional photons. These photons are clones of the incoming photon: the 16 photons are indistinguishable, they arrive on the detector within the same coherence time and spread on the same diffraction pattern (the pattern centered on the incoming direction of the astronomical photon). Importantly, the photons don't arrive on the same position on the detector: they are in the same mode, and their final position on the detector is thus determined by the same probability distribution. Each photon detection is however an independent random realization: the 16 photons spread over the diffraction pattern. When these 16 photons are detected within one coherence time -- that is typically within a femto- to nanosecond -- one knows that this corresponds to a burst generated by one incoming astronomical photon. One can thus keep the average of the 16 positions as the best estimate of the incoming direction of the astronomical photon. The precision on the incoming direction is then improved by a factor $\sqrt{16}=4$ with respect to the detection of a single photon, i.e. with respect to the diffraction limit. This gain in resolution could be increased by an arbitrarily large factor by choosing an arbitrarily large amplification gain. This however would violate the Heisenberg uncertainty principle. We have thus in a way re-demonstrated the no-cloning theorem in the context of astronomical imaging.

\subsection{Probabilistic amplification in the context of astronomical imaging}
Let us now benefit from the research done in quantum optics: it is possible to probabilistically amplify a photon with less than average noise. Crucially it is also possible to recognize low-noise amplification events. One may thus choose to select those events, while discarding  events with larger amplification noise. This would allow to gain in angular resolution while losing in sensitivity. 
To determine the limits of this trade-off, imagine that one could recognize how strongly a given photon had been diffracted as it passed the telescope aperture and arrived on the detector. One could then keep only those photons that were diffracted least. The point spread function would get thinner and one could distinguish finer features on the astronomical object, but the signal would decrease: to retrieve as many photons  as in the absence of selection, the exposure time would need to be increased. Figure 1 quantifies the trade-off: as the selection criterion becomes more severe and only the least diffracted photons are kept, the angular resolution improves and the sensitivity degrades.

   \begin{figure} [ht]
   \begin{center}
   \begin{tabular}{c} %% tabular useful for creating an array of images 
   \includegraphics[width=0.5\textwidth]{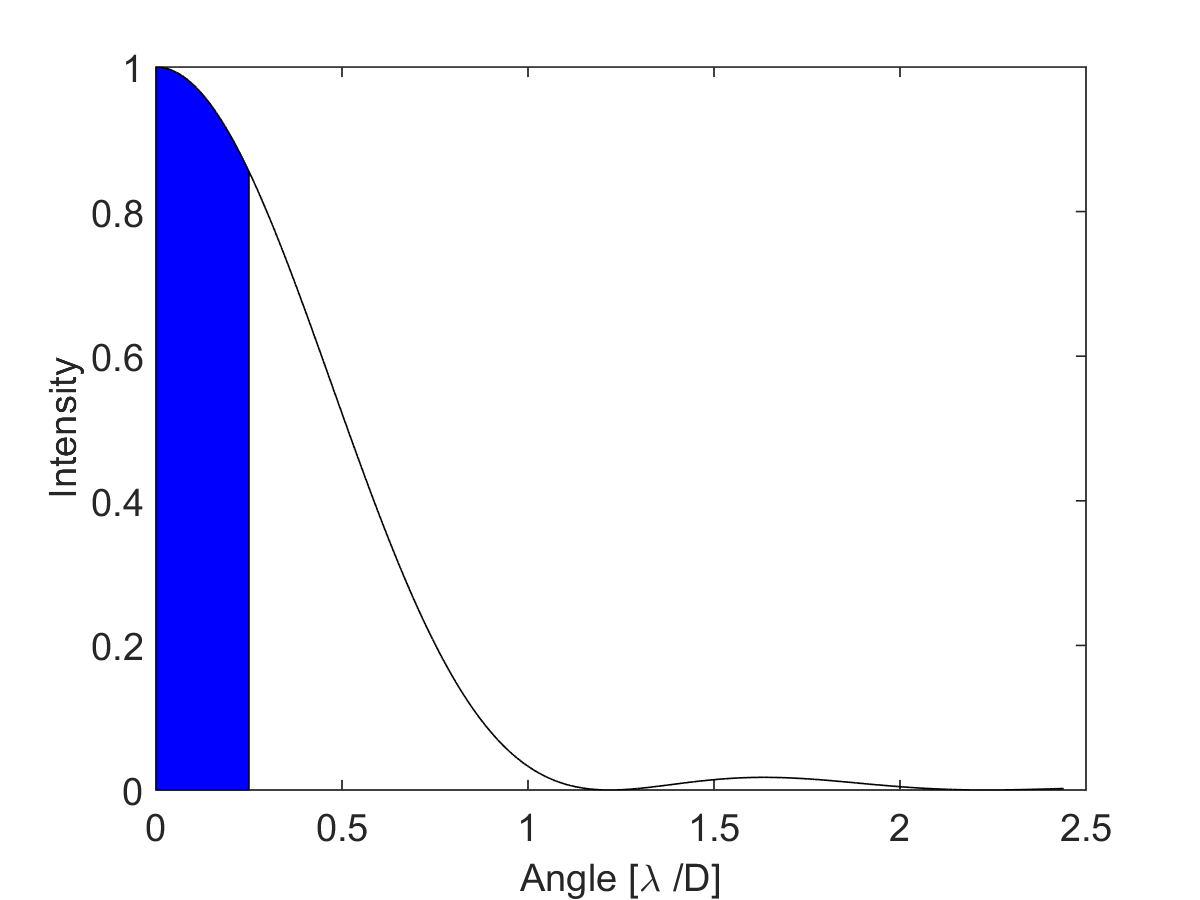}
   \includegraphics[width=0.5\textwidth]{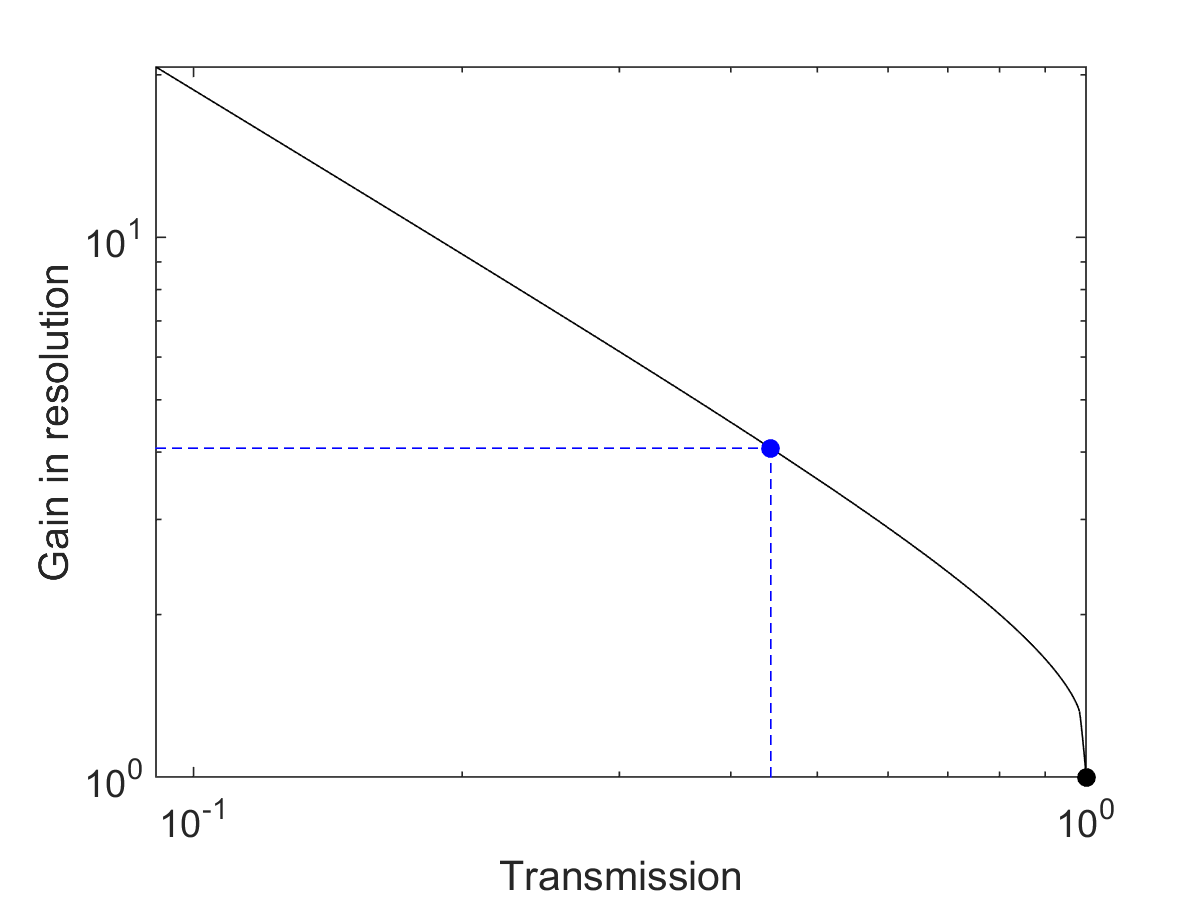}
   \end{tabular}
   \end{center}
   \caption[example] 
%>>>> use \label inside caption to get Fig. number with \ref{}
   { \label{fig:diffPattern} 
Left panel: Diffraction pattern and example of selection at $0.25\,\lambda/D$. Right panel: The y-axis equals $\sigma_0/\sigma$, where $\sigma$ (resp. $\sigma_0$) is the rms of the photon positions after (resp. without) selection\footnotemark: As one selects only the least diffracted photons, the rms of the photon positions decreases and the angular resolution increases. The blue point corresponds to the example on the left panel: the selection of photons diffracted at less than $0.25\,\lambda/D$. }
   \end{figure} 

\footnotetext{In absence of selection, the rms is calculated on the diffraction pattern up to radius $2.44\,\lambda/D$ (i.e. up to its second minimum) since the rms of the complete diffraction pattern through a circular aperture is infinite.}

Whatever the practical scheme -- probabilistic photon amplification or another approach -- the trade-off between gain in resolution and loss in sensitivity will be limited by Figure\,\ref{fig:diffPattern}: one can indeed not do better than accept the uncertainty principle, recognize by how much each photon was diffracted and keep only the least diffracted photons. 

As for the practical implementation of probabilistic single-photon amplification to astronomical imaging: in previous publications we have considered the possibility to send the incoming astronomical photon through a medium of excited atoms and to use stimulated emission as the amplification mechanism \cite{2014A&A...561A.118K, 2016OptL...41.3181K}. The noise then takes the form of spontaneous emissions. As mentioned here in introduction, an amplifier adds an average of $g$ noise photons per amplified mode (see for example \citenum{PhysRevD.26.1817}). Ideally the modes of the amplifier are adjusted to the telescope: each amplifier mode extends over a diffraction pattern and the coherence time of the photons emitted by the amplifier equal the coherence time of the astronomical photon, d$t=\lambda^2/(\Delta\lambda\,c)$. Under these optimal conditions, an amplifier of gain $g$ emits $g$ spontaneous photons per coherence time and per diffraction pattern. For a single input photon, the average number of amplified photons is then just equal to the average number of noise photons emitted in its mode. In addition the spatial and temporal distribution of the stimulated and spontaneous photons are identical: 
in both cases, a burst of photons arrives on the detector within a coherence time, d$t$, and spreads over a  diffraction pattern. This burst corresponds to a set of photons that have been stimulated by one initial photon : that initial photon was either the incident astronomical photon, or a spontaneously emitted photon. In both cases, the number of photons per burst follows a Bose-Einstein distribution of average and rms deviation equal to $g$, the amplifier gain. We have not yet found how stimulated photons could be distinguished from spontaneous photons, i.e. how low-noise amplification events could be recognized and selected. 

The schemes that are used in quantum optics rely on another type of amplification: see for example \citenum{haw2016surpassing}. Photons are prepared and mixed with the signal photon via beam-splitters. If the photons are in the same mode they interfere. This interference term allows to recognize a successful amplification event. In the absence of an interference term, the amplification event is discarded. These schemes could possibly be adapted for use in astronomical imaging (see \citenum{Marek}). 

\section{Conclusion}
In conclusion, probabilistic single photon amplification has been theoretically validated and practically demonstrated in quantum optical laboratories. Such a probabilistic amplification scheme should allow to overcome the diffraction limit of a telescope. Since the scheme is probabilistic, all incident astronomical photons can not be successfully amplified and one needs to discard a fraction of the incident photons. This article has indicated fundamental limits to the trade-off between gain in angular resolution and loss in sensitivity. The way to practically implement probabilistic single-photon amplification in astronomy remains an open question. 

\bibliographystyle{spiebib} % makes bibtex use spiebib.bst
\bibliography{report} % bibliography data in report.bib

\end{document}